\documentclass[manuscript]{acmart}

\usepackage{soul}
\usepackage{tfrupee}
\usepackage[normalem]{ulem}
\usepackage{natbib}

\usepackage{xcolor}
\definecolor{control}{RGB}{0, 76, 109}
\definecolor{intervention}{RGB}{105, 150, 179}
\definecolor{white}{RGB}{255, 255, 255}
\definecolor{black}{RGB}{0, 0, 0}
\definecolor{median}{RGB}{237, 157, 66}

\begin{document}


\title[Towards “Anytime, Anywhere” Community Learning and Engagement for Public Sector AI]{Towards “Anytime, Anywhere” Community Learning and Engagement around the Design of Public Sector AI}


\author{Wesley Hanwen Deng}
\affiliation{%
  \institution{Carnegie Mellon University}
  \city{Pittsburgh}
  \state{Pennsylvania}
  \country{USA}}
\email{hanwend@cs.cmu.edu}

\author{Motahhare Eslami}
\authornote{Both authors contributed equally to this research.}
\email{meslami@cs.cmu.edu}
\affiliation{%
  \institution{Carnegie Mellon University}
  \streetaddress{5000 Forbes Ave}
  \city{Pittsburgh}
  \state{PA}
  \postcode{15213}
  \country{USA}
}

\author{Kenneth Holstein}
\authornotemark[1]
\email{kjholste@cs.cmu.edu}
\affiliation{%
  \institution{Carnegie Mellon University}
  \streetaddress{5000 Forbes Ave}
  \city{Pittsburgh}
  \state{PA}
  \postcode{15213}
  \country{USA}
}

\renewcommand{\shortauthors}{TBA}

\begin{abstract}
Data-driven algorithmic and AI systems are increasingly being deployed to automate or augment decision processes across a wide range of public service settings. Yet community members are often unaware of the presence, operation, and impacts of these systems on their lives. With the shift towards algorithmic decision-making in public services, technology developers increasingly assume the role of de-facto policymakers, and opportunities for democratic participation are foreclosed. 
In this position paper, we articulate an early vision around the design of ubiquitous infrastructure for public learning and engagement around civic AI technologies.
Building on this vision, we provide a list of questions that we hope can prompt stimulating conversations among the HCI community.
\end{abstract}








\maketitle

\section{Background}

Data-driven algorithmic and AI systems are increasingly deployed to augment or automate public sector decision-making in high-stakes settings such as child welfare \cite{chouldechova2018case}, recidivism prediction \cite{chouldechova2017fair}, and public health care \cite{obermeyer2019dissecting}. Yet, community members are often unaware of the presence, operation, and impacts of these systems on their lives \cite{robertson2021modeling}. With this shift towards algorithmic decision-making, technology developers increasingly assume the role of de-facto policymakers \cite{alkhatib2019street}. Community members whose lives are directly impacted by these technologies are typically excluded from decision-making around their design and development. As a consequence, AI systems are often designed in ways that inadvertently amplify historical inequities, disproportionately harming the most marginalized members of our communities \cite{levy2021algorithms, mulligan2019procurement, veale2018fairness}. There is a great need to empower community members, especially those with lower technology literacy, to learn about and help to shape how AI technologies affect their communities \cite{boehner2016data, kuo2023understanding, shen2022model, robertson2021modeling}

Recent HCI research has focused on enhancing AI literacy and engagement around the design and oversight of civic AI technologies. For example, Long et al. co-designed a series of exhibits with community members aiming to enable informal learning experiences around AI technologies in public spaces like museums \cite{long2019designing, long2021co}. Lee et al. designed the ``WeBuildAI'' framework with the goal of enabling community members and relevant stakeholders with low AI literacy to ``build AI systems that represented their own beliefs,'' \cite{lee2019webuildai}. More recently, Alfrink et al. designed and implemented a framework for ``contestable AI'' framework as an initial step towards urban infrastructure that allows community members to contest the design and use of camera cars \cite{alfrink2022contestable, alfrink2022tensions}. However, \textbf{we lack methods that can support sustained, and continuous community learning and civic engagement} around public sector AI systems (cf. \cite{reynante2021framework, hsu2022empowering, le2016designing}). In addition, existing approaches often \textbf{fail to reach the most marginalized community members}, falling prey to broader challenges in fostering civic engagement \cite{long2019designing, le2016designing, reynante2021framework}. 

\section{Research Vision and Open Questions}
\begin{figure}[t]
  \centering
  \includegraphics[width=0.87\linewidth]{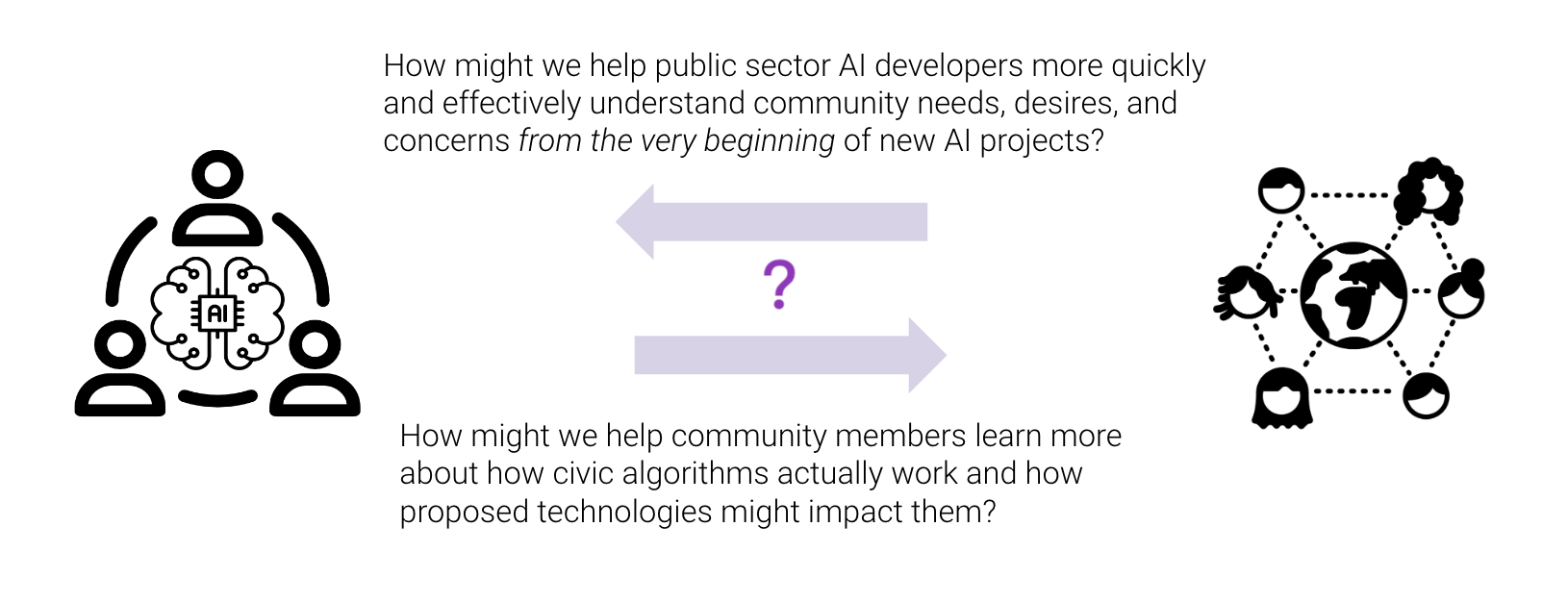}
  \caption{
    We envision the development of new infrastructure that can facilitate sustained bi-directional learning between local community members and public sector decision-makers \& technology designers.
}
  \label{fig}
\end{figure}

We envision a future in which diverse community voices are empowered to shape decisions around the design, development and oversight of public sector AI technologies that impact their communities. To this end, we invite the HCI community to collectively explore ways to support ``anytime, anywhere'' learning and engagement around the design of public sector AI technologies (Figure \ref{fig}). For example, how might public and community spaces (e.g., bus stops, parks, public libraries) be reimagined as sites for learning and civic engagement around the design of public sector technologies? Holding public conversations and events in accessible public places can help ensure that a wide range of people have the opportunity to participate and share their perspectives \cite{reynante2021framework}. It is particularly important to engage with those who may be disproportionately impacted by AI, including communities that have historically been underrepresented or marginalized in the technology design. 
How can we design to empower diverse community members, spanning a broad range of backgrounds and relevant literacies, to articulate their desires for new technologies that can better address their actual needs, as well as their concerns about technologies currently in use? How can we equip children and young adults with the informal learning opportunities and skills necessary for advocacy around public sector AI systems, including the ability to propose better alternatives to existing systems? As AI practitioners continue to recognize the importance of engaging users in the design and development of their AI systems \cite{deng2022understanding, deng2022exploring}, how might we develop tools and guidelines to facilitate meaningful collaboration between AI practitioners and community members?

By exploring such questions as a community, we hope to contribute towards a future in which government decision-makers, researchers, and technology developers recognize that, when properly empowered to do so, community members can be truly valuable collaborators in design and decision-making around public sector AI technologies. In our envisioned future, the excuse that community members are ``not technical enough'' to meaningfully engage around AI system design would be used less and less often as a justification against community involvement around impactful policy decisions that are disguised as purely “technical” decisions. We believe that achieving this vision requires innovation on local infrastructure for community learning and engagement, and that HCI researchers will have a critical role to play. 

\newpage

\section{Discussion}
In the workshop, we hope to open up conversations around how we might reimagine public spaces as environments for lifelong learning and civic engagement, and prepare community members to critically and constructively engage in a world where invisible, imperfect algorithms increasingly shape major aspects of their lives. In particular, we hope to spark discussion around the following three questions:

\begin{itemize}
    \item How might we empower \textbf{diverse} community members to engage as learners, co-designers, and overseers of AI technologies that are intended to benefit their communities?
    
    \item How can we provide opportunities for community members to engage at \textbf{a range of levels}, offering a “low floor” of brief, informal learning and design engagements, and a “high ceiling” of opportunities for longer-term civic engagement?
    
    \item How might we enable civic learning and engagement as \textbf{mutual, bi-directional} processes in which community members and public sector decision-makers and technology designers continuously communicate with and learn from each other?
\end{itemize}

\bibliographystyle{ACM-Reference-Format}
\bibliography{citation}

\end{document}